\documentclass[aps,pra,onecolumn,groupedaddress,showpacs]{revtex4}

\usepackage{graphics}
\usepackage{epsfig}
\usepackage{amsmath}

\begin{document}


\title{Two qubits entanglement dynamics in a symmetry-broken
environment}


\author{Marco Lucamarini$^{1}$}
\author{Simone Paganelli$^{1,2}$}
\author{Stefano Mancini$^{3}$}

\affiliation{\medskip {$^{1}$Dipartimento di Fisica,
Universit\`{a}
di Roma ``La Sapienza", I-00185 Roma, Italy.} \\
{$^{2}$Dipartimento di Fisica, Universit\`{a} di Bologna, via
Irnerio 46, I-40126 Bologna, Italy.} \\ {$^{3}$Dipartimento di
Fisica, Universit\`{a} di Camerino, I-62032 Camerino, Italy.}}


\date{\today}

\begin{abstract}
We study the temporal evolution of entanglement pertaining to two
qubits interacting with a thermal bath. In particular we consider
the simplest nontrivial spin bath models where symmetry breaking
occurs and treat them by mean field approximation. We analytically
find decoherence free entangled states as well as entangled states
with an exponential decay of the quantum correlation at finite
temperature.
\end{abstract}


\maketitle

\section{Introduction}

Since 1935 \cite{sch,epr}, entanglement has been recognized as one
of the most puzzling features of Quantum Mechanics. However, it is
nowadays a widespread opinion that it also represents a
fundamental resource for many quantum information protocols. As
such, entanglement deserves to be analyzed in all respects. A
primary concern is its robustness against environmental effects,
and a supplied literature exists aimed at preserving entanglement
coherence \cite{{vio1},{vio2},{vio3},{vio4},{vio5},{vio6},{vio7}}.
More recently, attention has been devoted to the problem of
\textit{thermal entanglement} \cite{{gun}} i.e.
quantifying entanglement arising in spin chains at thermal
equilibrium with a bath. In this approach environment determines
the temperature $T$ to allow for a thermal distribution of system
energy levels, while the detailed interaction between system and
environment is not an essential part of the matter. The same is
true also for those works that focus on entanglement decoherence
\cite{{yu},{yu2}} (also known as \textit{disentanglement}
\cite{{ter},{bla}}). In this context the study of entanglement
time behaviour is carried on with a master equation formalism and
markovian approximation \cite{gar} or, more generally, with
arguments provided by spin-boson models.

In the present paper we are going to envisage a novel approach to
the problem along the line introduced for the first time in
Ref.\cite{dep} (a similar outline but supported by numerical means
is also present in \cite{tess}). There, the authors considered a
one spin system interacting with a fermionic environment endowed
with a structure capable of symmetry-breaking \cite{sac}. It was shown
by
analytical methods that coherence time increases as magnetic order
enlarges or, in other terms, as temperature decreases. Here we
extend this argument to a two qubits system plunged in a fermionic
environment described by {\it Transverse Ising model} (TIM) and
{\it Ising model} (IM) \cite{sac}. We shall examine the time evolution
of
concurrence of the bipartite system \cite{woo}, and find
environment-limited concurrences as well as unlimited ones
according to environment ordering level.

The paper is organized as follows: in section II we introduce the
model by referring to \cite{dep} and we revise some results. In
section III we extend the model to a bipartite systems, and we
present the results of paradigmatic cases in Sec. IV. Finally,
Sec. V is for conclusions. Explicit calculations are reported in
Appendices A and B.

\section{The Model}

We consider the general scenario of a system and a bath described
by hamiltonians $H_{s}$ and $H_{B}$ respectively, and interacting
through the hamiltonian $H_{sB}$. The total hamiltonian is then
$H=H_{s}+H_{sB}+H_{B}$, and the initial density matrix is assumed
to be factorized, i.e., $\rho (0)=\rho_{s}\otimes\rho_{B}$. We are
looking for the time evolution of the reduced system density
matrix $\rho_{s}(t)$; in particular we are interested in its off
diagonal elements, the so called ``coherences''. If $H$ doesn't
depend on time the total density matrix will evolve accordingly to
\begin{equation}
\rho\left( t\right) =e^{-iHt}\rho(0)e^{iHt}\,.\label{rho}
\end{equation}
We can then obtain the reduced density matrix by tracing out the
bath degrees of freedom in Eq.(\ref{rho})
\begin{equation}
\rho_{s}\left( t\right) =tr_{B}\rho(t)\,.\label{rhoS}
\end{equation}

We now follow the line sketched in Ref.\cite{dep} to introduce the
model for a spin system interacting with a spin bath. First of
all, we assume the bath density matrix having a thermal
distribution, that is $\rho_{B}= (e^{-H_{B}/T})/Z$, with $T$ the
bath temperature multiplied by the Boltzmann constant, and
$Z=tr\left( e^{-H_{B}/T}\right)$ the partition function.
Furthermore, we ask the bath hamiltonian to be a ``symmetry
breakable'' one, that is endowed with phase transition in the
degrees of freedom that provide the coupling with the system. The
simplest hamiltonian with these requirements is a long ranged
\textit{Ising Model}-like one (IM). We add to it a transverse
field to include a more general case in the analysis, dealing
eventually with a \textit{Transverse Ising Model} bath hamiltonian
(TIM). The differences between the two models are minimal as
coherence and entanglement is concerning and, in any case, we will
be able to find results for IM in the limit of no transverse field
for TIM. These peculiar environment hamiltonians will be studied
through mean field approximation \cite{sac}.

\subsection{TIM-environment}

\label{TIs1}

Let us consider $N+1$ spin-$\frac{1}{2}$, and let $S_{j}^{\alpha}$
be the $\alpha$ component ($\alpha=x,\,y,\,z$) of the $j$th spin
($j=0,1,\ldots,N$). The label $j=0$ refers to the system operators
while $j=1,\ldots,N$ to the bath operators. Furthermore,
$S_{j}^{\pm}=(S_{j}^{x}\pm iS_{j}^{y})/2$ are the spin flip
operators, and $|0\rangle$ and $|1\rangle$ are the lower and upper
eigenstates of $S^{z}$. The following hamiltonians define the
energy of the system, of the TIM-bath and of the interaction
between them:
\begin{subequations}
\label{TrIsHam}
\begin{eqnarray}
H_{s} &=&-\mu_{0}\,S_{0}^{z},\\
H_{sB} &=&-\frac{J_{0}}{\sqrt{N}}S_{0}^{z}\sum\limits_{k}S_{k}^{z}\,,\\
H_{B}
&=&-w\sum\limits_{k}S_{k}^{x}-\frac{J}{N}\sum\limits_{i,k}S_{i}
^{z}S_{k}^{z}\,,
\end{eqnarray}
\end{subequations}
where $\mu_{0}$ is the coupling constant with an external magnetic
field parallel to the $\hat{z}$ axis, $J_{0}$, $J$ are exchange
coupling constants and $w$ is the strength of the transverse
field; they are all non negative constants. The indices of the
sums run from $1$ to $N$. Eq.(\ref{TrIsHam}c) describes a material
in which spins compete to align along the positive direction of
$\hat{x}$ axis or along $\hat{z}$ axis following a ferromagnetic
behaviour; of course in the latter case the absolute direction of
alignment is not important since the hamiltonian is symmetric in
$z$-operators. We can notice that energy exchanges between system
and bath are not included in the interaction hamiltonian; this
will generate a pure dephasing dynamics, in which energy will be
conserved, and temporal evolution analytically solved.

The main difficulty with Eqs.(\ref{TrIsHam}) is represented by the
nonlinear term in $H_{B}$. For this reason it is helpful to
approximate it with a mean field bath hamiltonian, as explained in
\cite{dep}:
\begin{equation}
H_{B}^{mf}=-w\sum\limits_{k}S_{k}^{x}
-2Jm\sum\limits_{k}S_{k}^{z}+m^{2}JN\,.
\label{MFTrIsHam}
\end{equation}
In the above equation $m$ is the order parameter of the phase
transition. Its absolute value ranges from $0$ to $\frac{1}{2}$ as
long as temperature ranges from the critical value $T_{c} =
\frac{J}{2}$ to $0$ : the greater $|m|$ the larger the magnetic
order of the bath along $\hat{z}$ axis. In the following we are
going to consider only positive values for $m$ since results are
sign-independent. Everything remains true with the substitution $m
\rightarrow -m$. This is a consequence of $H_{B}$ $z$-symmetry,
that is not lost in $H_{B}^{mf}$. The order parameter $m$ is
implicitly defined by the following self-consistent equation for
the quantity $\Theta=\pm\sqrt{w^{2}+4m^{2}J^{2}}$ (also $\Theta$'s
sign, written here for sake of precision, is irrelevant, for the
same reasons of $m$'s):
\begin{equation}
\frac{\Theta}{J}=\tanh\frac{\Theta}{2T}.\,\label{ScEq2}
\end{equation}
It is worth noting that from Eq.(\ref{ScEq2}) we have
$\Theta\rightarrow J$ for $T\rightarrow0$; furthermore, from the
definition of $\Theta$, we can see it tends to $2mJ$ in the limit
of no transverse field ($w\rightarrow0$).

Together with Eq.(\ref{ScEq2}) we must consider the following
condition on the transverse field to obtain an ordered phase with
TIM:
\begin{equation}
\frac{w}{J}<\tanh\left(\frac{w}{2T}\right).\label{ScEqW}
\end{equation}
This condition is not satisfied in the range of temperatures above
$T_{c}$; for this reason the whole formalism we are using is valid
only in the broken phase.

With the linearized mean field bath hamiltonian it is possible to
evaluate the coherence of the system (see Appendix \ref{appB}):
\begin{eqnarray}
S_{0}^{-}(t)&=&tr_{B}\left[ e^{-iH^{mf}t}\left( S_{0}
^{-}(0)\otimes\rho_{B}\right) e^{iH^{mf}t}\right] \nonumber\\
&=&\frac{1}{Z}tr_{B}\left[ e^{-iH^{mf}t}\left( \left|
0\right\rangle \left\langle 1\right| \otimes
e^{-H_{B}^{mf}/T}\right) e^{iH^{mf}t} \right]
\nonumber\\
&=&S_{0}^{-}(0)r_{_{TIM}}(t)\label{Coher1}
\end{eqnarray}
where $H^{mf}=H_{s}+H_{sB}+H^{mf}_{B}$, and
\begin{equation}\label{rTIM}
r_{_{TIM}}(t)=\left[ \cos\left( \frac
{tmJJ_{0}}{\Theta\sqrt{N}}\right) +i\frac{\Theta}{J}\sin\left(
\frac{tmJJ_{0}}{\Theta\sqrt{N}}\right) \right].
\end{equation}
Equation (\ref{Coher1}) tells us that the time evolution of the
off diagonal term of the system density matrix, responsible for
the coherence of the system, is enclosed in the time behaviour of
the complex valued factor $r_{_{TIM}}(t)$. In particular, in order
to find system decoherence, we ask whether and when this factor's
absolute value goes to zero. In the limit of large $N$ we can
approximate it as:
\begin{equation}
\left| r_{_{TIM}}(t)\right| \approx\exp\left[ {-\frac{J_{0}^{2}
m^{2}t^{2}}{2}\left( \frac{J^{2}}{\Theta^{2}}-1\right) }\right]
\,.\label{r2}
\end{equation}
We can see from (\ref{r2}) that the system coherence decays
exponentially with time. The coherence time is:
\begin{equation}
\tau_{_{TIM}}=\frac{\left| \Theta
\right|}{J_{0}m}\sqrt{\frac{2}{J^{2}-\Theta^{2}}} \,,
\label{time1}
\end{equation}
and increases as temperature decreases; for $T=0$ it is
$\tau=\infty$, and the system remains coherent. This is quite a
counter-intuitive effect since collective quantum properties of
materials endowed with phase transition disappear as ordering
increases (see for instance Ref. \cite{jon}). The factor $t^{2}$
in the exponent denotes the intrinsically reversible nature of the
process, in contrast to irreversibility introduced by markovian
approach, and is closely related to the ``Zeno effect''
\cite{zur}. In particular the periodicity of $r_{_{TIM}}(t)$ in
Eq.(\ref{rTIM}) leads to the so called ``recoherences'' on a
Poincar\'{e} time scale. Decoherence takes place in the limit of
an environment with infinite degrees of freedom; besides, the same
limit is necessary to support the mean field theory approach we
adopted. Thus in this context the limit $N\rightarrow\infty$ has a
double function: to take into account the decoherence process
and to give a meaning to the mean field approximation written above.

We briefly notice here that the factor $r_{_{TIM}}(t)$ in
Eq.(\ref{r2}) is exactly alike to $r_{as}(t)$ of Eq.(32) in
\cite{dep}. But as far as that paper is concerning we must point
out some inaccuracies: the final result (32) is correct, but the
intermediate steps to find it are not. In particular the general
formula (11) applies only if the $3\times3$ matrices $\Lambda_{k}$
commute, and this is not true when you look at Eq.(29) of that
article. For this reason the intermediate formula (30) is wrong
and the oscillations showed in Fig.1 are not present.

\subsection{Limit of no transverse field: IM-environment}


 In the limit of $w\rightarrow0$ we obtain from (\ref{TrIsHam}) the
 IM-hamiltonians which lead to
\begin{equation}
\left| r_{_{IM}}(t)\right| \approx\exp\left[
-\frac{J_{0}^{2}t^{2}} {2}\left( \frac{1}{4}-m^{2}\right) \right]
\,.\label{r1}
\end{equation}
We can notice the same behaviour as for TIM-bath, but slightly
more transparent: the coherence time is
$\tau_{_{IM}}=\frac{2}{J_{0}}\sqrt{\frac{2}{1-4m^{2}}}$ and its
limits are $\tau_{_{IM}}^{(T=T_{c})}=\frac{2\sqrt{2}}{J_{0}}$ and
$\tau_{_{IM}}^{(T=0)}=\infty$. We note that coherence explicit
dependence on bath coupling constant $J$ has disappeared in this
case; only interaction coupling constant $J_{0}$ enters coherence
expression when the bath is an IM-one. Otherwise the $J$ coupling
is indirectly present in (\ref{r1}) because it has a role in
determining the order parameter $m$ by means of Eq.(\ref{ScEq2}).

\section{The Extension}

In this section we extend results obtained in the previous one by
considering a two qubits system, and studying the time evolution
of their entanglement. We assume that the system qubits, labeled
by $01$ and $02$, interact between them and with environment, that
is symmetry-breakable and modeled by TIM hamiltonians generalizing
those of Eqs.(\ref{TrIsHam}):
\begin{subequations}
\label{HamTrIs2}
\begin{eqnarray}
H_{s} &=&-\xi_{0}S_{01}^{z}S_{02}^{z},\\
H_{sB}&=&-\frac{J_{0}}{\sqrt{N}}\left(S_{01}^{z}+S_{02}^{z}\right)
\sum\limits_{k}S_{k}^{z}\,,\\
H_{B}&=&-w\sum\limits_{k}S_{k}^{x}-\frac{J}{N}
\sum\limits_{i,k}S_{i}^{z}S_{k}^{z}\,,
\end{eqnarray}
\end{subequations}
In above equations $\xi_{0}$ represents the coupling constant
between the qubits. We have discarded both local interactions,
like that between qubits and an external magnetic field, and local
couplings with environment degrees of freedom, a situation
resembling a ``collective'' system-environment pairing
\cite{vio5}.

As a measure of entanglement between two qubits we adopt the so
called ``concurrence'' \cite{woo}, which ranges from $0$ for
separable states to $1$ for maximally entangled states. The
concurrence is given by:
\begin{equation} C=\max\left\{
\lambda_{1}-\lambda_{2}-\lambda_{3}-\lambda _{4},0\right\}
\text{,}\label{Conc}
\end{equation}
where $\lambda_{1}$, $\lambda_{2}$, $\lambda_{3}$ and
$\lambda_{4}$ are the square roots of the eigenvalues, in
decreasing order, of the matrix $R=\rho_{s}\widetilde{\rho_{s}}$.
Here $\rho_{s}$ is the density matrix of the 2 system qubits, and
$\widetilde{\rho_{s}}$ is the ``time reversed'' matrix given by
\begin{equation}
\widetilde{\rho_{s}}=\left( \sigma^{y}_{01}\otimes\sigma ^{y}_{02}
\right) \rho_{s}^{\ast}\left( \sigma^{y}_{01}\otimes\sigma
^{y}_{02} \right) \,,
\end{equation}
where $\sigma$'s are the usual Pauli matrices. The symbol
$\rho_{s}^{\ast}$ means complex conjugation of the matrix
$\rho_{s}$ in the standard basis $\left| 00\right\rangle $,
$\left| 01\right\rangle $, $\left| 10\right\rangle $, $\left|
11\right\rangle $.

We assume that the qubits are initially decoupled from the
environment, and the
bath having a thermal density matrix $\rho_{B}=(e^{-H_{B}/T})/Z$.
Therefore, we can write the whole state as:
\begin{equation}
\rho=\left| \Psi\right\rangle \left\langle \Psi\right|
\otimes\rho_{B}\label{rho2}
\end{equation}
with a generic system pure state:
\begin{equation}
\left| \Psi\right\rangle =\alpha\left| 00\right\rangle
+\beta\left| 01\right\rangle +\gamma\left| 10\right\rangle
+\delta\left| 11\right\rangle
\,,\qquad|\alpha|^{2}+|\beta|^{2}+|\gamma|^{2}+|\delta
|^{2}=1\,.\label{states1}
\end{equation}
The steps to find time evolution of Eq.(\ref{rho2}) are similar to
those leading to Eq.(\ref{Coher1}) (see Appendix \ref{appB}), but
now operators are represented by $4\times4$ matrices, being our
system composed by two qubits. After mean field approximation
(\ref{MFTrIsHam}) for the bath hamiltonian and some elementary
algebra we obtain the reduced density matrix as:
\begin{equation}
\rho_{s}\left( t\right) =tr_{B}\left( \rho\left( t\right)
\right) =\left(
\begin{array}{cccc}
\left| \alpha\right| ^{2} & \alpha^{\ast}\beta
A^{\ast}e^{-\frac{1}{2}it\xi_{0}} & \alpha^{\ast}\gamma
A^{\ast}e^{-\frac{1}{2}it\xi_{0}} & \alpha^{\ast}\delta
B^{\ast}\\
\alpha\beta^{\ast}Ae^{\frac{1}{2}it\xi_{0}} & \left| \beta\right|
^{2} &
\beta^{\ast}\gamma & \beta^{\ast}\delta
A^{\ast}e^{\frac{1}{2}it\xi_{0}}\\
\alpha\gamma^{\ast}Ae^{\frac{1}{2}it\xi_{0}} & \beta\gamma^{\ast}
& \left|
\gamma\right| ^{2} & \gamma^{\ast}\delta
A^{\ast}e^{\frac{1}{2}it\xi_{0}}\\
\alpha\delta^{\ast}B &
\beta\delta^{\ast}Ae^{-\frac{1}{2}it\xi_{0}} & \gamma
\delta^{\ast}Ae^{-\frac{1}{2}it\xi_{0}} & \left| \delta\right|
^{2}
\end{array}
\right) \,,
\end{equation}
where the coefficients
\begin{subequations}
\label{ABD}
\begin{eqnarray}
A &=&\left[ \cos\left( \frac{tmJJ_{0}}{\Theta\sqrt{N}}\right)
+i\frac{ \Theta }{J}\sin\left(
\frac{tmJJ_{0}}{\Theta\sqrt{N}}\right) \right]
^{N}\,,\\
B &=&\left[ \cos\left( \frac{2tmJJ_{0}}{\Theta\sqrt{N}}\right)
+i\frac{\Theta}{J}\sin\left(
\frac{2tmJJ_{0}}{\Theta\sqrt{N}}\right) \right] ^{N}\,,
\end{eqnarray}
\end{subequations}
characterize the time dependence of the concurrence. From the
above expression of $\rho_{s}(t)$ we can find the matrix $R(t)$
and its eigenvalues, and from them, as explained, the final
concurrence of the system. The complete expression for $R(t)$ and
for coefficients of Eq.(\ref{ABD}) is given in Appendix
\ref{appC}.
In the following we are going to consider
some paradigmatic cases for the initial state (\ref{rho2}).

\section{Paradigmatic cases}

\subsubsection{Case 1}

Let us set $\alpha=\delta=0$ in Eq.(\ref{states1}) for the initial
state of the system. We obtain $\left| \Psi\right\rangle
=\beta\left| 01\right\rangle +\gamma\left| 10\right\rangle $ and
$R$ matrix reduces to:
\begin{equation}
R(t)=\left(
\begin{array}{cccc}
0 & 0 & 0 & 0\\
0 & 2\left| \beta\right| ^{2}\left| \gamma\right| ^{2} &
2\beta^{\ast
}\left| \beta\right| ^{2}\gamma & 0\\
0 & 2\beta\gamma^{\ast}\left| \gamma\right| ^{2} & 2\left|
\beta\right|
^{2}\left| \gamma\right| ^{2} & 0\\
0 & 0 & 0 & 0
\end{array}
\right) \,,
\end{equation}
whose square rooted eigenvalues are:
\begin{subequations}
\begin{eqnarray}
\lambda_{1} &=& 2\left| \beta\right| \left| \gamma\right| \,,\\
\lambda_{2} &=&\lambda_{3}=\lambda_{4}=0\,,
\end{eqnarray}
\end{subequations}
This leads to the following concurrence:
\begin{equation}
\label{CTIM1}
C_{_{TIM}}=2\left| \beta\right| \left|
\gamma\right| \,.
\end{equation}
The entanglement results time independent, so the state does not
perceive the presence of the environment. The reason is that
$|\Psi\rangle$ is an eigenstate of the interaction hamiltonian and
so it represents a decoherence free entangled state \cite{vio5}.
Since $w$ is not present in the concurrence written above we know
that the expression for the concurrence would be exactly the same
for an IM-environment.

\subsubsection{Case 2}

Now we set $\beta=\gamma=0$ in Eq.(\ref{states1}) and obtain the
state $\left| \Psi \right\rangle =\alpha\left| 00\right\rangle
+\delta\left| 11\right\rangle $. The $R$ matrix becomes:
\begin{equation}
R(t)=\left(
\begin{array}{cccc}
\left| \alpha\right| ^{2}\left| \delta\right| ^{2}\left(
1+\left| B\right| ^{2}\right) & 0 & 0 & 2\alpha^{\ast}\left|
\alpha\right|
^{2}\delta B^{\ast}\\
0 & 0 & 0 & 0\\
0 & 0 & 0 & 0\\
2\alpha\delta^{\ast}\left| \delta\right| ^{2}B & 0 & 0 & \left|
\alpha\right| ^{2}\left| \delta\right| ^{2}\left( 1+\left|
B\right| ^{2}\right)
\end{array}
\right) \,,
\end{equation}
with square rooted eigenvalues in decreasing order:
\begin{subequations}
\label{sre}
\begin{eqnarray}
\lambda_{1} &=&\left| \alpha\right| \left| \delta\right|
\left( \left|
B\right| +1\right) \,,\\
\lambda_{2} &=&\left| \alpha\right| \left| \delta\right|\left(
\left| \left|
B\right| -1\right|\right) \,,\\
\lambda_{3} &=&\lambda_{4}=0\,.
\end{eqnarray}
\end{subequations}
>From Eqs.(\ref{ABD}), for large $N$, we get:
\begin{equation}
\left| B\right| \approx\exp\left[ -2J_{0}^{2}m^{2}t^{2}\left(
\frac{J^{2}}{\Theta^{2}}-1\right) \right]. \
\end{equation}
Then, by using concurrence definition and Eqs.(\ref{sre}), we
arrive at:
\begin{equation}
\label{CTIM2} C_{_{TIM}}=2\left| \alpha\right| \left|
\delta\right| \left| B\right| =2\left| \alpha\right| \left|
\delta\right| \exp\left[ -2J_{0}^{2} m^{2}t^{2}\left(
\frac{J^{2}}{\Theta^{2}}-1\right) \right]. \
\end{equation}
\begin{figure}
\begin{center}
\includegraphics[width=0.5\textwidth]{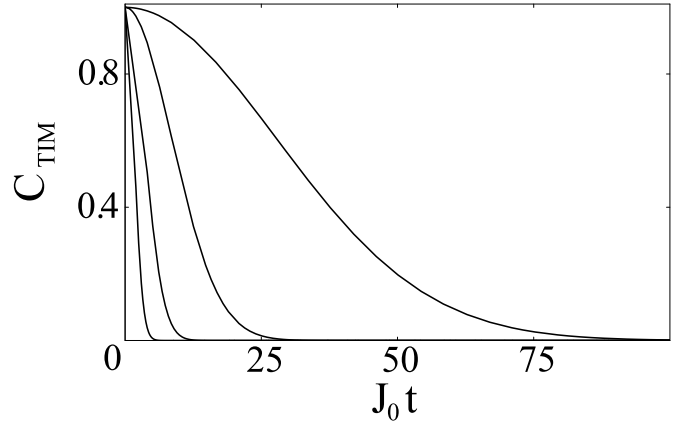}
\end{center}
\caption{\label{CFig1}
Concurrence versus scaled time $J_{0}t$.
Curves from the left to the right are for
$\frac{T}{T_{c}}= \{.75, .50, .35, .25\} $.
The value of other parameter are
$w=0.1, J=2$.}
\end{figure}
The time behaviour of the concurrence just obtained is shown in
Fig.\ref{CFig1} for different values of the ratio
$\frac{T}{T_{c}}$. We notice that in this case the qubits perceive
the presence of the thermal bath, which spoils entanglement
between them; in fact the initial state is no longer an eigenstate
of the interaction hamiltonian. Only for zero temperature the
order parameter reaches its saturation value and the concurrence
remains constant. The behaviour is very similar to that of one
qubit system coherence described by Eq.(\ref{Coher1}), but
entanglement decoherence is exactly twice faster than one qubit
decoherence. This result agrees with what found in \cite{yu2}.
Furthermore, together with the previous case, it falls
within the general limitations represented by the {\it
Universal Disentangling Machine} \cite{ter}.

In the limit $w\rightarrow0$ we obtain the concurrence for an
IM-bath:
\begin{equation}
\label{C12} C_{_{IM}}=2\left| \alpha\right| \left| \delta\right|
\exp\left[ -2J_{0}^{2}t^{2}\left( \frac{1}{4}-m^{2}\right) \right]
\,.
\end{equation}
Analogously to what already noticed for the single qubit
coherence, in this limit the factor $J$ disappears from the
explicit concurrence expression. The only exchange coupling
constant that enters in the decoherence time for the concurrence
is $J_{0}$.

\subsubsection{Case 3}

If we set $\alpha=\beta=0$ we obtain a product state $\left|
\Psi\right\rangle =\gamma\left| 10\right\rangle +\delta\left|
11\right\rangle =\left( \gamma\left| 0\right\rangle +\delta\left|
1\right\rangle \right) \left| 1\right\rangle $, which trivially
gives:
\begin{equation}
R(t)=\left( \mathbf{0}\right) \Longrightarrow C=0\,.
\end{equation}
In this case TIM hamiltonians are not able to induce entanglement
between system qubits.

\subsubsection{Case 4}

\begin{figure}
\begin{center}
\includegraphics[width=0.5\textwidth] {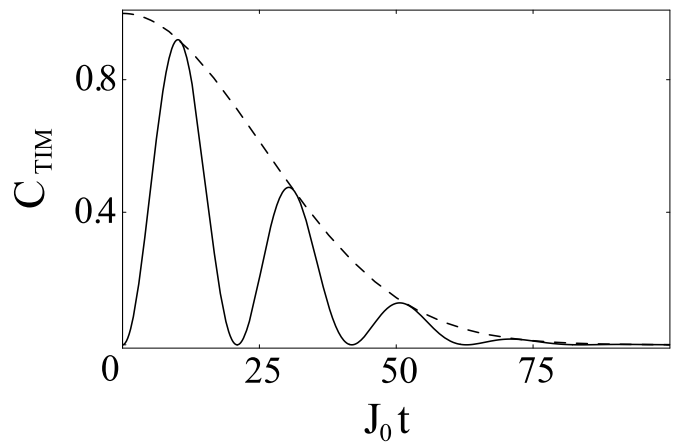}
\end{center}
\caption{\label{CFig2} Concurrence versus the scaled time
$J_{0}t$. The plot shows the limiting role of decoherence (dotted
line) that falls down exponentially, on entanglement (continuous
line). The value of parameters are $w=0.1, \xi_{0}=0.3, J=2,
\frac{T}{T_{c}}=.25$. }
\end{figure}

If we set $\alpha=\beta=\gamma=\delta=\frac{1}{2}$ we obtain again
a separable initial state, but different from the previous one:
$\left| \Psi\right\rangle =\frac{1}{2}\left( \left|
00\right\rangle +\left| 01\right\rangle +\left| 10\right\rangle
+\left| 11\right\rangle \right) =\frac{1}{\sqrt{2}}\left( \left|
0\right\rangle +\left| 1\right\rangle \right)
\frac{1}{\sqrt{2}}\left( \left| 0\right\rangle +\left|
1\right\rangle \right) $. In this case the $R$ matrix is not
trivial:
\begin{equation}
R(t)=\frac{1}{16}\left(
\begin{array}{cccc}
1+\left| B\right| ^{2}-2\left| A\right| ^{2}e^{-it\xi_{0}} &
U_{\xi_{0}} & U_{\xi_{0}}%
& 2B^{\ast}-2\left( A^{\ast}\right) ^{2}e^{-it\xi_{0}}\\
-V_{\xi_{0}} & 2-2\left| A\right| ^{2}e^{it\xi_{0}} & 2-2\left|
A\right|
^{2}e^{it\xi_{0}} & -U_{\xi_{0}}\\
-V_{\xi_{0}} & 2-2\left| A\right| ^{2}e^{it\xi_{0}} & 2-2\left|
A\right|
^{2}e^{it\xi_{0}} & -U_{\xi_{0}}\\
2B-2A^{2}e^{-it\xi_{0}} & V_{\xi_{0}} & V_{\xi_{0}} & 1+\left|
B\right| ^{2}-2\left| A\right| ^{2}e^{-it\xi_{0}}
\end{array}
\allowbreak\right) \,,
\end{equation}
where:
\begin{subequations}
\label{UV}
\begin{eqnarray}
U_{\xi_{0}} &=&\left( 2A^{\ast}e^{-\frac{1}{2}it\xi_{0}}-\left(
A^{\ast}+AB^{\ast
}\right) e^{\frac{1}{2}it\xi_{0}}\right) \\
V_{\xi_{0}} &=&\left( 2Ae^{-\frac{1}{2}it\xi_{0}}-\left(
A+A^{\ast}B\right) e^{\frac{1}{2}it\xi_{0}}\right)
\end{eqnarray}
\end{subequations}
The concurrence is evaluable esplicitly, but the expression
is too much cumbersome therefore not reported here.
We only show in
Fig.\ref{CFig2} its behaviour.
The concurrence starts from its null value and increases because of
the interaction between system qubits. If there wasn't
disentanglement it would reach its maximum and decrease again
giving rise to oscillations of equal amplitude.
Nevertheless, the presence of
environment alters this temporal behaviour
damping the oscillations.
For suitable values
of coupling constants it can even prevent qubits from entangling
at all. The interesting question of the maximal entanglement
generation under dephasing processes arises naturally in this case
\cite{yu}.

\section{Conclusion}

We have studied time behaviour of entanglement between two qubits
dipped in a large symmetry-breakable fermionic environment, below
the critical temperature $T_{c}$. In the frame of mean field
theory analytical results are provided for concurrence of the
bipartite system, with temperature as a parameter of the problem.
The hamiltonians involved in the discussion are those typical of
\textit{Transverse Ising Models} (TIM), capable of magnetic
ordering under suitable conditions. To assign them a physical
meaning we notice that, upon addition of a transverse field in
$H_{s}$, our model resembles an array of Rydberg atoms interacting
with a cavity mode of the radiation field \cite{dep}. Nevertheless
such an assumption for $H_{s}$ makes the problem unsolvable by
analytical techniques, and requires numerical investigation that
we plan to accomplish in a near future. Beside that an important
improvement would be to overcome mean field approximations adopted
in the text, by including the effect of fluctuations, or by
applying the {\it spin wave} approach \cite{spa} to the bath.

What comes out from the paper is quite a counterintuitive
conservation of entanglement in a bath with strong interactions:
the bigger the coupling strength (or the lower the ratio
$\frac{T}{T_{c}}$) the longer the time qubits remain entangled
(Eq.(\ref{CTIM2}) and Fig.(\ref{CFig1})). In some cases entangled
qubits don't perceive environment at all, and the system state is
a \textit{decoherence free} one (Eq.(\ref{CTIM1})). Several
connections with results from the field of entanglement
decoherence are provided. We believe our analysis can be useful to
complete knowledge about entanglement dynamical properties.

\section*{Acknowledgements}


\appendix

\section{}\label{appB}

\subsection{Exponentiation of suitable matrices}

Let us define a $2 \times 2$ traceless matrix ${\cal A}$ as
\begin{equation}
{\cal A}=\left( a\sigma _{x}+b\sigma _{z}\right) =\left(
\begin{array}{cc}
b & a \\
a & -b
\end{array}
\right)\,,
\end{equation}
with $a$, $b$ real coefficients. The exponentiation of $A$ gives:
\begin{equation}
\label{expA} e^{\cal A}=\left( \cosh q\right) I+\left( \frac{\sinh
q}{q}\right) {\cal A} \,; e^{i\cal A}=\left( \cos q\right)
I+i\left( \frac{\sin q}{q}\right) {\cal A}
\end{equation}
with $q=\sqrt{\left( a^{2}+b^{2}\right) }$. Therefore:
\begin{equation}
\label{A} tr\left( e^{\cal A}\right) =2\left( \cosh q\right) \,;
tr\left( e^{i\cal A}\right) =2\left( \cos q\right)
\end{equation}
Let us extend these arguments to three matrices ${\cal I}$, ${\cal
R}$, and ${\cal I}^{\prime }$ of the same form of ${\cal A}$:
\begin{eqnarray}\label{3matr}
tr\left[ e^{i\cal I}e^{\cal R}e^{{i\cal I}^{\prime }}\right]
&=&tr\left\{ \left[ \left( \cos x\right) I+i\left( \frac{\sin
x}{x}\right) {\cal I}\right] \left[ \left( \cosh y\right) I+\left(
\frac{\sinh y}{y}\right) {\cal R}\right] \left[ \left( \cos
z\right) I+i\left( \frac{\sin z}{z}\right) {\cal I}^{\prime }
\right] \right\} \nonumber \\
&=&\left( \cosh y\right) \left[ 2\left( \cos x\right) \left( \cos
z\right) +i\left( \cos z\right) \left( \frac{\sin x}{x}\right)
\left( \frac{\tanh y}{y}\right) tr\left({\cal I}{\cal R}\right)
\right.\nonumber \\
&&\qquad\qquad\left.+i\left( \cos x\right) \left( \frac{\tanh
y}{y}\right) \left( \frac{\sin z}{z} \right) tr\left({\cal R}{\cal
I}^{\prime }\right) -\left( \frac{\sin x}{x}\right) \left(
\frac{\sin z}{z} \right) tr\left({\cal I}{\cal I}^{\prime }\right)
\right]\,,
\end{eqnarray}
where $x$, $y$, $z$ are respectively related to the elements of
${\cal I}$, ${\cal R}$, ${\cal I}'$ as $q$ was related to $A$.

\subsection{Coherence Expression for TIM}

As an example of calculation we report the steps that lead to
Eq.(\ref{r2}). All other calculations are easier than this one and
can be performed following the same line.

The time evolution of the total density matrix is:
\begin{eqnarray}
\rho \left( t\right) &=& \frac{e^{-m^{2}\tilde{J}N/T}}{Z}
\left\{
\exp\left\{it\sum\limits_{k}\left[
\left( \frac{J_{0}}{\sqrt{N}}S_{0}^{z}+2mJ\right)
S_{k}^{z}+wS_{k}^{x}\right] \right\}\rho _{s}
\right.\nonumber\\
&&\left.\times
\exp\left\{(1/T)\sum\limits_{k}\left(
wS_{k}^{x}+2mJS_{k}^{z}\right) \right\}
\exp\left\{-it\sum\limits_{k}\left[ \left( \frac{
J_{0}}{\sqrt{N}}S_{0}^{z}+2mJ\right) S_{k}^{z}+wS_{k}^{x}\right]
\right\}\right\}\,.
\label{rhoTI}
\end{eqnarray}

First, the partition function results:
\begin{equation}
Z=e^{-m^{2}\tilde{J}N/T}tr\left\{ \exp\left[
(1/T) \sum\limits_{k}\left(
wS_{k}^{x}+2mJS_{k}^{z}\right)\right]\right\}
=e^{-m^{2}\tilde{J}
N/T}\prod\limits_{k}tr\left[ e^{\left( wS_{k}^{x}+2m
JS_{k}^{z}\right)/T }\right]\,.
\end{equation}
By virtue of equation (\ref{A}) we find
\begin{equation}
Z = e^{-m^{2}\tilde{J}N/T}\;2^{N}\;\left\{ \cosh \left[ \frac{\Theta
}{2T}
\right] ^{N}\right\}\,.
\end{equation}
Notice that the constant $e^{-m^{2}\tilde{J}N/T}$ in the
partition function simplifies with that present in Eq.(\ref{rhoTI}).

Let us now study the time evolution of the operator
$S_{0}^{-}=\left| 0\right\rangle \left\langle 1\right| $ that
represents the off diagonal part of the density matrix:
\begin{eqnarray}\label{S0tr3}
S_{0}^{-}(t)&=&
\left[ 2\cosh \left( \frac{\Theta}{2T} \right) \right]^{-N}
tr_{B}
\left\{ \prod\limits_{k}e^{it [ ( \frac{J_{0}}{
\sqrt{N}}S_{0}^{z}+2mJ ) S_{k}^{z}+wS_{k}^{x} ] }e^{(
wS_{k}^{x}+2mJS_{k}^{z})/T }\left| 0\right\rangle \left\langle
1\right| \prod\limits_{k}e^{-it [ ( \frac{J_{0}}{\sqrt{N}}
S_{0}^{z}+2mJ ) S_{k}^{z}+wS_{k}^{x} ] }\right\}
\nonumber\\
&=& S_{0}^{-}(0) \left[ 2\cosh \left( \frac{\Theta }{2T} \right)
\right]^{-N} \prod\limits_{k}tr_{B} \left\{ e^{i\cal I}e^{\cal
R}e^{{i\cal I}^{\prime}}\right\} \,,
\end{eqnarray}
where:
\begin{subequations}
\begin{eqnarray}
{\cal I} &=&t\left[ \left( \frac{J_{0}}{2\sqrt{N}}+2mJ\right)
S_{k}^{z}+wS_{k}^{x}\right]\,, \\
{\cal R} &=&\left(wS_{k}^{x}+2mJS_{k}^{z}\right)/T \,, \\
{\cal I}^{\prime } &=&-t\left[ \left(
-\frac{J_{0}}{2\sqrt{N}}+2mJ\right) S_{k}^{z}+wS_{k}^{x}\right]\,.
\end{eqnarray}
\end{subequations}
In order to use Eq.(\ref{3matr}) we evaluate the following
quantities:
\begin{subequations}
\begin{eqnarray}
x&=&\frac{t}{2}\sqrt{\Theta
^{2}+2\frac{mJJ_{0}}{\sqrt{N}}+O\left(\frac{1}{N}\right)} \\
y&=&\frac{
\Theta }{2T}\Longrightarrow \left( \frac{\tanh y}{y}\right) =\frac{2T}{
J} \\
z&=&\frac{t}{2}\sqrt{\Theta ^{2}-2\frac{mJJ_{0}}{
\sqrt{N}}+O\left(\frac{1}{N}\right)}
\end{eqnarray}
\end{subequations}
and
\begin{subequations}
\begin{eqnarray}
tr\left({\cal I}{\cal R}\right) &=& \frac{t}{2T}\left(
\frac{mJJ_{0}}{\sqrt{N}}+\Theta ^{2}\right)\,,
\\
tr\left({\cal R}{\cal I}^{\prime }\right) &=& \frac{t}{2T}\left(
\frac{mJJ_{0}}{\sqrt{N}}-\Theta^{2}\right)\,,
\\
tr\left({\cal I}{\cal I}^{\prime }\right) &=&
-\frac{t^{2}}{2}\left( \Theta
^{2}-\frac{1}{4}\frac{J_{0}^{2}}{N}\right) =-\frac{t^{2}\Theta
^{2}}{2}+O\left(\frac{1}{N}\right)\,.
\end{eqnarray}
\end{subequations}
Then, substituting these into Eq.(\ref{3matr}) and performing the
product we obtain:
\begin{equation}
\prod\limits_{k}tr\left\{
e^{i\mathcal{I}}e^{\mathcal{R}}e^{i\mathcal{I} ^{\prime }}\right\}
=2^{N}\left( \cosh \frac{\Theta }{2T}\right) ^{N}\left[ \cos
\left( \frac{tmJJ_{0}}{\Theta \sqrt{N}}\right) +i\frac{\Theta
}{J}\sin \left( \frac{tmJJ_{0}}{\Theta \sqrt{N}}\right) \right]
^{N}\,. \label{prodtr3}
\end{equation}
We can recognize in the second member of Eq.(\ref{prodtr3}) the
constant $r_{_{TIM}}(t)$ defined in Eq.(\ref{rTIM}); the absolute
value of it, in the limit of large $N$, gives the result of
Eq.(\ref{r2}). The other quantities of the article come out with
similar calculations.

\section{}\label{appC}

\subsection*{Complete $R$ matrix for TIM}

Let's begin with time dependent density matrix expression for TIM
hamiltonians (\ref{HamTrIs2}). After mean field approximation
(\ref{MFTrIsHam}) we obtain:
\begin{eqnarray}\label{tr3}
\rho\left(t\right)&=&\frac{1}{Z}\left[e^{-it
\left(H_{s}+H_{sB}+H_{B}^{mf}\right)}\rho
_{s}e^{-H_{B}^{mf}/T}e^{it\left(H_{s}+H_{sB}+H_{B}^{mf}\right)}\right]
\nonumber \\
&=& \frac{1}{Z}\exp \left\{ it\left[ \sum\limits_{k}\left(
\frac{J_{0}}{\sqrt{N}}\left(S_{01}^{z}+S_{02} ^{z}\right)
+2mJ\right)S_{k}^{z}+\sum\limits_{k}wS_{k}^{x}\right]\right\} \rho
_{s}^{\prime }\exp \left\{ (1/T)\sum\limits_{k}\left(wS_{k}
^{x}+2mJS_{k}^{z}\right) \right\}\nonumber \\
&&\qquad\qquad\left. \exp\left\{ -it\left[
\sum\limits_{k}\left(\frac{J_{0}}{\sqrt{N}}\left(
S_{01}^{z}+S_{02}^{z}\right)+2mJ\right)S_{k}^{z}+\sum\limits_{k}wS_{k}
^{x}\right]\right\}\right.\,,
\end{eqnarray}
Where we've have set $\rho _{s}^{\prime }=e^{it\xi
_{0}S_{01}^{z}S_{02}^{z}}\rho _{s}e^{-it\xi
_{0}S_{01}^{z}S_{02}^{z}}$.

The constants present in Eqs.(\ref{ABD}) are found by complex
conjugation of the following quantities, evaluated in a similar
manner as the one seen in Appendix \ref{appB}:
\begin{subequations}
\label{ABDStar}
\begin{eqnarray}
A^{\ast } &=&\frac{1}{Z}\prod\limits_{k}tr_{B}\left\{
e^{it[2mJS_{k}^{z}+wS_{k}^{x}]}e^{(wS_{k}^{x}+2mJS_{k}^{z})/T}
e^{-it[(\frac{
J_{0}}{\sqrt{N}}+2mJ)S_{k}^{z}+wS_{k}^{x}]}\right\} \\
B^{\ast } &=&\frac{1}{Z}\prod\limits_{k}tr_{B}\left\{
e^{it[(-\frac{J_{0}}{
\sqrt{N}}+2mJ)S_{k}^{z}+wS_{k}^{x}]}
e^{(wS_{k}^{x}+2mJS_{k}^{z})/T}e^{-it[(
\frac{J_{0}}{\sqrt{N}}+2mJ)S_{k}^{z}+wS_{k}^{x}]}\right\} \\
D^{\ast } &=&\frac{1}{Z}\prod\limits_{k}tr_{B}\left\{
e^{it[(-\frac{J_{0}}{ \sqrt{N}}
+2mJ)S_{k}^{z}+wS_{k}^{x}]}e^{(wS_{k}^{x}+2mJS_{k}^{z})/T}
e^{-it[2mJS_{k}^{z}+wS_{k}^{x}]}\right\}
\end{eqnarray}
\end{subequations}
After calculations it's an easy task to verify that $A^{\ast
}=D^{\ast }$, and for this reason the constant $D$ doesn't appear
in Eqs.(\ref{ABD}).

The matrix $R(t)$ for TIM is:
\begin{equation}
\label{Rmatr2}
R(t) = \left(
\begin{array}{cc}
R_{1} & R_{2} \\
R_{3} & R_{4}
\end{array}
\right)
\end{equation}
\begin{equation}
R_{1}=\left(
\begin{array}{cc}
\left| \alpha \right| ^{2}\left| \delta \right| ^{2}\left(
1+\left| B\right| ^{2}\right) -2\alpha ^{\ast }\beta \gamma \delta
^{\ast }\left| A\right| ^{2}e^{-it\xi_{0} } & 2\alpha ^{\ast
}\beta \left| \gamma \right| ^{2}A^{\ast }e^{-\frac{1}{2}it\xi_{0}
}-\left| \alpha \right| ^{2}\gamma ^{\ast }\delta
\left( A^{\ast }+AB^{\ast }\right) e^{\frac{1}{2}it\xi_{0} } \\
\alpha \beta ^{\ast }\left| \delta \right| ^{2}\left( A+A^{\ast
}B\right) e^{ \frac{1}{2}it\xi_{0} }-2\left| \beta \right|
^{2}\gamma \delta ^{\ast }Ae^{- \frac{1}{2}it\xi_{0} } & -2\alpha
\beta ^{\ast }\gamma ^{\ast }\delta \left| A\right|
^{2}e^{it\xi_{0} }+2\left| \beta \right| ^{2}\left| \gamma \right|
^{2}
\end{array}
\right)
\end{equation}

\begin{equation}
R_{2}=\left(
\begin{array}{cc}
2\alpha^{\ast}\left| \beta\right| ^{2}\gamma
A^{\ast}e^{-\frac{1}{2}it\xi_{0} }-\left| \alpha\right|
^{2}\beta^{\ast}\delta\left( A^{\ast}+AB^{\ast }\right)
e^{\frac{1}{2}it\xi_{0}} & 2\alpha^{\ast}\left| \alpha\right|
^{2}\delta B^{\ast}-2\left( \alpha^{\ast}\right)
^{2}\beta\gamma\left(
A^{\ast}\right) ^{2}e^{-it\xi_{0}} \\
-2\alpha\left( \beta^{\ast}\right) ^{2}\delta\left| A\right|
^{2}e^{it\xi_{0} }+2\beta^{\ast}\left| \beta\right| ^{2}\gamma &
\left| \alpha\right| ^{2}\beta^{\ast}\delta\left(
A^{\ast}+AB^{\ast}\right) e^{\frac{1}{2}it\xi_{0}
}-2\alpha^{\ast}\left| \beta\right| ^{2}\gamma
A^{\ast}e^{-\frac{1}{2}it\xi_{0}}
\end{array}
\right)
\end{equation}

\begin{equation}
R_{3}=\left(
\begin{array}{cc}
\alpha\gamma^{\ast}\left| \delta\right| ^{2}\left(
A+A^{\ast}B^{\ast }\right) e^{\frac{1}{2}it\xi_{0}}-2\beta\left|
\gamma\right| ^{2}\delta^{\ast }Ae^{-\frac{1}{2}it\xi_{0}} &
-2\alpha\left( \gamma^{\ast}\right) ^{2}\delta\left| A\right|
^{2}e^{it\xi_{0}}+2\beta\gamma^{\ast}\left|
\gamma\right| ^{2} \\
2\alpha\delta^{\ast}\left| \delta\right| ^{2}B-2\beta\gamma\left(
\delta^{\ast}\right) ^{2}A^{2}e^{-it\xi_{0}} & 2\beta\left|
\gamma\right|
^{2}\delta^{\ast}Ae^{-\frac{1}{2}it\xi_{0}}-\alpha\gamma^{\ast}\left|
\delta\right| ^{2}\left( A+A^{\ast}B\right)
e^{\frac{1}{2}it\xi_{0}}
\end{array}
\right)
\end{equation}
\begin{equation}
R_{4}=\left(
\begin{array}{cc}
-2\alpha\beta^{\ast}\gamma^{\ast}\delta\left| A\right|
^{2}e^{it\xi_{0} }+2\left| \beta\right| ^{2}\left| \gamma\right|
^{2} & \left| \alpha\right|
^{2}\gamma^{\ast}\delta\left( A^{\ast}+AB^{\ast}\right) e^{\frac{1}{2}%
it\xi_{0}}-2\alpha^{\ast}\beta\left| \gamma\right| ^{2}A^{\ast
}e^{-\frac{1}{2}%
it\xi_{0}} \\
2\left| \beta\right| ^{2}\gamma\delta
Ae^{-\frac{1}{2}it\xi_{0}}-\alpha \beta^{\ast}\left| \delta\right|
^{2}\left( A+A^{\ast}B\right) e^{\frac {1}{2}it\xi_{0}} & \left|
\alpha\right| ^{2}\left| \delta\right| ^{2}\left( 1+\left|
B\right| ^{2}\right) -2\alpha^{\ast}\beta\gamma\delta^{\ast
}\left| A\right| ^{2}e^{-it\xi_{0}}
\end{array}
\right)
\end{equation}
From it we have extracted all particular cases treated in the
text.

\end{document}